\documentclass[aps,pra,preprint]{revtex4-1}
\usepackage[colorlinks=true, linkcolor=blue, citecolor=blue, urlcolor=blue]{hyperref}
\usepackage{natbib}
\usepackage{graphicx} 
\usepackage{subfigure}
\usepackage{bm}
\usepackage{amssymb,amsmath}

\begin{document}

\title{Band Structure and Dynamics of Single Photons in Atomic Lattices}
\author{Wenxuan Xie}
\affiliation{Department of Applied Physics, Yale University, New Haven, CT 06511, USA}
\email{wenxuan.xie@yale.edu}
\author{John C. Schotland}
\affiliation{Department of Mathematics and Department of Physics, Yale University, New Haven, CT 06511, USA} 
\email{john.schotland@yale.edu}

\begin{abstract}
 We present a framework to investigate the collective properties of atomic lattices in one, two, and three dimensions. We analyze the single-photon band structure and associated atomic decay rates, revealing a fundamental dependence on dimensionality. One- and two-dimensional arrays are shown to be inherently radiative, exhibiting band gaps and decay rates that oscillate between superradiant and subradiant regimes, as a function of lattice spacing. In contrast, three-dimensional lattices are found to be fundamentally non-radiative due to the inhibition of spontaneous emission, with decay only at discrete Bragg resonances. Furthermore, we demonstrate that this structural difference dictates the system dynamics, which crosses over from dissipative decay in lower dimensions to coherent transport in three dimensions. Our results provide insight into cooperative effects in atomic arrays at the single-photon level.
\end{abstract}
  
\maketitle

%%===================================================%%
%%           Introduction                            %%
%%===================================================%%
\section{\label{sec:intro}INTRODUCTION}
Periodic arrays of quantum emitters, realized with both natural and artificial atoms, constitute a powerful platform for exploring cooperative light-matter interactions \cite{deutsch1995photonic,van1996photonic,de1998point,zeng2006complete,klugkist2006mode,antezza2009fano,antezza2009spectrum,bienaime2012controlled,antezza2013quantum,bettles2016enhanced,guerin2016subradiance,syzranov2016emergent,yoo2016cooperative,asenjo2017atom,asenjo2017exponential,bettles2017topological,perczel2017topological,perczel2020topological,PhysRevA.96.063801,plankensteiner2017cavity,shahmoon2017cooperative,guimond2019subradiant,javanainen2019light,zhang2019theory,ballantine2020subradiance,bettles2020quantum,parmee2020signatures,williamson2020superatom,PhysRevA.104.063707,rubies2022photon,PhysRevA.108.030101,PhysRevLett.132.163602,PhysRevA.111.053712}. In these systems, often termed atomic lattices, emitters are typically modeled as stationary sites interacting via photon exchange. These photon-mediated dipole-dipole interactions are inherently radiative and long-ranged, generating a rich landscape of collective phenomena governed by lattice geometry. Fundamental among these are superradiance and subradiance, which arise from the constructive or destructive interference of emitted radiation and serve as the basis for diverse collective optical effects \cite{dicke1954coherence,skribanowitz1973observation,gross1982superradiance}.

The behavior of atomic lattices is dimension dependent. In one-dimensional (1D) systems, for instance in waveguide quantum electrodynamics, photon confinement mediates strong atom-atom coupling, allowing for the investigation of optical nonlinearities \cite{shen2007strongly,astafiev2010resonance,lodahl2017chiral,RevModPhys.95.015002,chang2014quantum}. Two-dimensional (2D) lattices have attracted significant attention for exhibiting phenomena such as enhanced reflection \cite{bettles2016enhanced,shahmoon2017cooperative,srakaew2023subwavelength}, subradiant Bell states \cite{guimond2019subradiant}, and topological edge states \cite{bettles2017topological,perczel2017topological,perczel2020topological}. Moreover, 2D arrays are utilized in quantum simulation, facilitating the study of exotic phases including Mott insulators, topological spin liquids, and 2D antiferromagnets \cite{browaeys2020many,ebadi2021quantum,daley2022practical,maskara2025programmable,bakr2010probing,semeghini2021probing,scholl2021quantum}. Early theoretical work on three-dimensional (3D) lattices established methods for calculating band structures, focusing on photonic band gaps and quantum fluctuations \cite{van1996photonic,de1998point,zeng2006complete,klugkist2006mode,antezza2009fano,antezza2009spectrum}. Unlike lower-dimensional systems, where photons can escape in at least one direction, 3D lattices confine light in all directions, leading to the inhibition of spontaneous emission \cite{yablonovitch1987inhibited,blanco1998cds,ogawa2004control,lodahl2004controlling,vion2009manipulating,PhysRevLett.107.143902,schulz2024strongly}.

The theoretical description of these problems—particularly band structure and dynamical properties—requires an accurate evaluation of photon-mediated coupling. This task is computationally demanding because it involves complex-valued, long-ranged lattice sums that converge slowly and are often only conditionally convergent. Previous studies have developed several techniques to evaluate such lattice sums and the associated band structures for 2D and 3D arrays \cite{antezza2009fano,antezza2009spectrum,perczel2017topological,PhysRevA.96.063801,perczel2020topological}. A widely used strategy in these works is to transform the lattice sum into momentum space via the Poisson summation formula. However, this procedure necessitates the introduction of a regularization cutoff to control the attendant divergences. The necessity for this cutoff is twofold. First, the photon-mediated interactions decay too slowly to permit a straightforward application of the Poisson summation formula. Second, and more fundamentally, the transformation to momentum space includes an on-site self-energy term. Since the self-energy corresponds to a divergent single-atom Lamb shift, the entire sum must be regularized to render a finite result. Although this cutoff is often interpreted as a physical manifestation of quantum fluctuations that effectively ``spread out'' the divergence \cite{perczel2017topological,PhysRevA.96.063801,perczel2020topological}, it inevitably introduces an artificial length scale into the theory.

In this paper, we develop a unified regularization-free framework for computing lattice sums in all dimensions. In contrast to existing momentum-space approaches, our method evaluates lattice sums in real space. This allows for the removal of the self-energy term without introducing an artificial cutoff. The resulting lattice sum is shown to converge by using a combination of the theta-function transform and the Ewald summation technique \cite{borwein2013lattice,ewald1921berechnung}. This approach yields rapidly convergent and compact expressions for the band structure, collective decay rates, and dynamical properties of atomic lattices in any dimension.
We apply this formalism to infinite atomic lattices in one, two, and three dimensions. For 1D and 2D lattices, we obtain  band structures with complex energies, in agreement with previous studies \cite{asenjo2017atom,asenjo2017exponential}. In contrast, the 3D lattice exhibits a purely real band structure, confirming that an infinite 3D array is non-radiative and suppresses spontaneous emission \cite{yablonovitch1987inhibited}. An analysis of collective decay rates in 1D and 2D reveals oscillatory behavior as a function of the lattice spacing, with rates approaching the single-atom value at large separations, consistent with recent numerical results \cite{sierra2022dicke,rui2020subradiant}. Our dynamical analysis further shows that excitations in 1D and 2D lattices eventually decay, whereas in 3D they persist in time without decay.

The remainder of this paper is organized as follows. Section~\ref{sec:model} introduces the model and derives the formula determining the single-photon band structure. Section~\ref{sec:1d} presents the band structure, decay rates, and dynamics of a one-dimensional lattice. Sec.~\ref{sec:2d} extends the analysis to two-dimensional square lattices. In Sec.~\ref{sec:3d}, we examine three-dimensional simple-cubic lattices and demonstrate a non-radiative band structure with Bragg resonances. Our conclusions are presented in Sec.~\ref{sec:discussion}. Additional technical details are provided in the Appendices.

%%===================================================%%
\section{\label{sec:model}MODEL}
We consider a $d$-dimensional lattice of two-level atoms interacting with the quantized electromagnetic field in three-dimensional free space. The atoms are positioned at lattice sites and sufficiently far apart to neglect short-range interatomic interactions. While we focus on square and simple-cubic geometries for 2D and 3D systems, respectively, and employ a scalar field model for simplicity, our methods generalize readily to the full vector electromagnetic field and other lattice structures.

The Hamiltonian of the system is of the form
\begin{equation}
 \hat H =   \hbar \Omega \sum\limits_{j} \hat{\sigma}^{\dagger}_j \hat{\sigma}_j + \sum\limits_{\mathbf{k}}~ \hbar \omega_{\mathbf{k}} \hat{a}_{\mathbf{k}}^{\dagger} \hat{a}_{\mathbf{k}} + \hbar g \sum\limits_{j} \sum\limits_{\mathbf{k}}~ ({e}^{\mathrm{i} \mathbf{k} \cdot \mathbf{r}_j}\hat{\sigma}^{\dagger}_j \hat{a}_{\mathbf{k}} + {e}^{-\mathrm{i} \mathbf{k} \cdot \mathbf{r}_j} \hat{\sigma}_j \hat{a}_{\mathbf{k}}^{\dagger}) .
    \label{eq:hamiltonian}
\end{equation}
Here we have imposed the usual rotating wave and dipole approximations. In addition,
%$\Omega=ck_{0}$ 
$\Omega$ is the atomic resonance frequency, $\omega_{\mathbf{k}}=c\left\vert \mathbf{k} \right\vert$ is the photon frequency, and $\hat{a}_{\mathbf{k}}^{\dagger}$ ($\hat{a}_{\mathbf{k}}$) creates (annihilates) a field mode with wavevector $\mathbf{k}$. The atomic raising (lowering) operators for the $j$th atom located at the points $\mathbf{r}_j$ are $\hat{\sigma}^{\dagger}_j$ ($\hat{\sigma}_j$). The coupling strength $g$ is assumed to be frequency-independent, which is a standard approximation in scalar quantum optics \cite{scully1997quantum}. 

We restrict our analysis to the single-excitation subspace (consistent with the rotating wave approximation) that is spanned by states with either one excited atom or one photon. The most general single-excitation state is defined by
\begin{equation}
    \lvert \Psi(t) \rangle  = \left( \sum\limits_{j} \psi_{j}(t) \hat{\sigma}_{j}^{\dagger} + \sum\limits_{\mathbf{k}}c_{\mathbf{k}} (t) \hat{a}_{\mathbf{k}}^{\dagger}  \right) \lvert 0 \rangle ,
    \label{eq:def of one photon state}
\end{equation}
where $\psi_{j}(t)$ and $c_{\mathbf{k}} (t)$ are the amplitudes for exciting the $j$th atom and creating a photon in the field mode with wavevector $\mathbf{k}$, respectively at time $t$ and $\lvert 0 \rangle$ denotes the combined ground state of the atoms and the field. The state 
obeys the Schrödinger equation 
\begin{equation}
\mathrm{i}\hbar \frac{\partial}{\partial t } \lvert \Psi(t) \rangle = \hat H  \lvert \Psi(t) \rangle , 
\end{equation}
from which it follows that $\psi_{j}(t)$ and $c_{\mathbf{k}}(t)$ evolve according to 
\begin{subequations}
    \begin{align}
     \mathrm{i} \frac{\partial }{\partial t} \psi_{j}(t) &= \Omega \psi_{j}(t) + g \sum \limits_{\mathbf{k}}  {e}^{\mathrm{i} \mathbf{k} \cdot \mathbf{r}_{j}}c_{\mathbf{k}} (t), \\
     \mathrm{i} \frac{\partial }{\partial t} c_{\mathbf{k}} (t) &= \omega_{\mathbf{k}}c_{\mathbf{k}}(t) + g \sum \limits_{j} {e}^{-\mathrm{i} \mathbf{k} \cdot \mathbf{r}_{j}} \psi_{j}(t).
    \end{align}
    \label{eq:eqs of motion}
\end{subequations}
We assume that initially there is a single excitation in the atomic system and a photon is not present in the field. We thus impose the initial conditions $\psi_{j}(0) \neq 0$ and $c_{\mathbf{k}}(0) = 0$. Taking the Laplace transform in time of \eqref{eq:eqs of motion} and applying the initial conditions yields, after eliminating $c_{\mathbf{k}}(s)$, a closed set of equations for the atomic amplitudes $\psi_{j}(s)$:
\begin{equation}
    (\mathrm{i} s -\Omega) \psi_{j}(s)  + \frac{g^{2} V}{ c } \sum\limits_{l} G(\mathbf{r}_{j}-\mathbf{r}_{l}; \frac{ \mathrm{i} s }{ c }) \psi_{l}(s) = \mathrm{i} \psi_{j}(0) .
    \label{eq:equation only with psi}
\end{equation}
Here we have defined the Laplace transform by ${f}(s)=\int_0^{\infty} f(t) {e}^{-s t} d t$ and replaced the summation over modes by an integral according to
$\sum_{\mathbf{k}} \to V/(2\pi)^{3} \int d^{3} {k}$. In addition, $G(\mathbf{r};k)$ is the Green's function for the nonlocal operator $\sqrt{-\Delta} - k$, which obeys the radiation condition \cite{hiltunen2024nonlocal}:
\begin{equation}
    \begin{aligned}
    G(\mathbf{r};k) = &\frac{1}{(2\pi)^{3}} \int d^{3} {q} \frac{{e}^{ \mathrm{i} \mathbf{q} \cdot \mathbf{r}}}{\left\vert  \mathbf{q}\right\vert -k} = \frac{1}{2 \pi^2 r^{2}}-\frac{ \mathrm{i} k}{4 \pi^2r }\left({e}^{\mathrm{i} kr} \displaystyle{E}_1( \mathrm{i} kr)\right.\left.-{e}^{- \mathrm{i} kr} \displaystyle{E}_1(-\mathrm{i} kr)\right)+\frac{k {e}^{\mathrm{i} kr}}{2 \pi r} ,
    \end{aligned}
     \label{eq:result of Green's function}
\end{equation}
where $\displaystyle{E}_{1}(z)$ denotes the exponential integral, which is defined by
\begin{equation}
{\displaystyle E}_{1}(z)=\int _{z}^{\infty }{\frac {e^{-t}}{t}}\,dt,\qquad |{\rm {Arg}}(z)|<\pi  .
\end{equation}
We note that $G$ is complex-valued, rendering the effective Hamiltonian for the atomic system non-Hermitian.
%The first two terms describe real dipole-dipole interactions, while the final imaginary term accounts for radiative exchange, rendering the effective Hamiltonian non-Hermitian. 
%%%
% write the effective Hamiltonian
%%%
The derivation of Eq.~\eqref{eq:result of Green's function} is provided in Appendix~\ref{appd:Greens function}.

The atoms are taken to belong to a $d$-dimensional lattice $\Lambda$ with lattice spacing $a$.
% atomic positions are $\mathbf{r}_{j} = a \sum_{\alpha=1}^{d} n_{j,\alpha} \hat{e}_{\alpha}$. 
In view of the translational invariance of the lattice, we take the Fourier transform of both sides of Eq.~\eqref{eq:equation only with psi} thus obtaining
\begin{align}
  \mathcal{H}(\mathbf{q},s) \psi(\mathbf{q},s) &= \mathrm{i} \psi(\mathbf{q},0),
    \label{eq:solution in q space}
\end{align}
where 
\begin{equation}
    \mathcal{H}(\mathbf{q},s) =  \mathrm{i} s -\Omega + \frac{g^{2} V}{ c } \sum_{\mathbf{r}\in \Lambda} G\left(\mathbf{r}; \frac{ \mathrm{i} s }{ c }\right) e^{-\mathrm{i}\mathbf{q}\cdot \mathbf{r}} .
\label{eq:def_H}
\end{equation}
Here the lattice Fourier transform is defined by $f(\mathbf{q}) = \sum_{\mathbf{r}\in\Lambda} {e}^{-\mathrm{i} \mathbf{q} \cdot \mathbf{r}}f(\mathbf{r})$, where $\mathbf q$ belongs to the first Brillouin zone (FBZ) of the lattice $\Lambda$.
%\begin{align}
%    \psi(\mathbf{q},s) &= \mathrm{i} \mathcal{H}^{-1} (\mathbf{q},s)\psi(\mathbf{q},0),
%    \label{eq:solution in q space}
%\end{align}
%where 
%\begin{equation}
%    \mathcal{H}(\mathbf{q},s) =  \mathrm{i} s -\Omega + \frac{g^{2} V}{ c } \sum_{\mathbf{r}_{j}} G(\mathbf{r}_{j}; \frac{ \mathrm{i} s }{ c }) {e}^{-\mathrm{i}\mathbf{q}\cdot \mathbf{r}_{j}}.
%\end{equation}
%
% INVERSION OF LAPLACE TRANSFORM
%

The dynamics of the system are obtained by inversion of the Laplace transform. 
We thus obtain
\begin{align}
\label{eq:inv_laplace}
\nonumber
\psi(\mathbf{q},t) &= \frac{1}{2\pi\mathrm{i}}\int_C e^{st} \psi(\mathbf{q},s) ds \\
&= \frac{1}{2\pi}\int_C  \frac{e^{st}}{\mathcal{H}\mathbf{q},s)} \psi(\mathbf{q},0) ds
\end{align}
where we have used Eq.~\eqref{eq:solution in q space} and the contour of integration C is parallel to the imaginary axis in the complex $s$-plane, lying to the right of any singularities of the integrand.
It follows that the singularities of the integrand, which correspond to the roots of the equation $\mathcal{H}(\mathbf{q},s) = 0$ govern the dynamics. It follows that it is necessary to find the roots of the equation
\begin{equation}
    \alpha -\alpha_{0} + 2 \pi \mathrm{i} \kappa \alpha^{2} + \kappa \mathrm{S}^{(d)}(\alpha, \bm{\beta}) = 0 ,
    \label{eq: pole equation}
\end{equation}
where the dimensionless quantities $\alpha = \mathrm{i} s a / 2\pi c$, $\alpha_{0} = \Omega a / 2 \pi c$, $\bm{\beta} = \mathbf{q} a / 2 \pi$ and $\kappa = g^{2} V / 2 \pi a c^{2}$. In order to obtain this result, we have separated out the $\mathbf{r}=0$ (self-energy) term in the sum appearing in Eq.~\eqref{eq:def_H} and have introduced the lattice sum
\begin{equation}
    \mathrm{S}^{(d)}(\alpha, \bm{\beta}) = \sum\limits_{\mathbf{n} \in \mathbb{Z}^{d} \backslash \{\mathbf{0}\}} G( \mathbf{n}; 2 \pi \alpha) {e}^{-2 \pi \mathrm{i}  \bm{\beta} \cdot \mathbf{n} }.
    \label{eq: def of lattice sum}
\end{equation}
In Eq.~\eqref{eq: pole equation}, the divergent real part of the self-energy term (Lamb shift) is absorbed into $\Omega$ while its finite imaginary part corresponds to the single-atom spontaneous emission rate. 
%
% WE SHOULD SAY MORE HERE
%

The solutions $\alpha(\bm{\beta})$ to Eq.~\eqref{eq: pole equation} define the band structure of the system. The real part and imaginary parts of $\alpha(\bm{\beta})$ account for the collective energy shift and the decay rate, respectively. Computing the inverse Laplace and Fourier transforms of Eq.~\eqref{eq:solution in q space} then yields the full  dynamics of the system.

%%===================================================%%
%%           One dimension                           %%
%%===================================================%%
\section{\label{sec:1d}One Dimensional Lattice}

%%===================================================%%

\subsection{\label{subsec:1d band structure}Band Structure and Pole Approximation}
\begin{figure}
   \centering
    \subfigure{\includegraphics[width=0.45\textwidth]{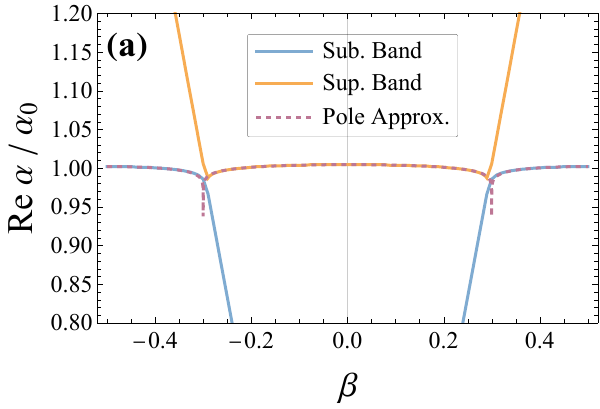}}
    \subfigure{\includegraphics[width=0.45\textwidth]{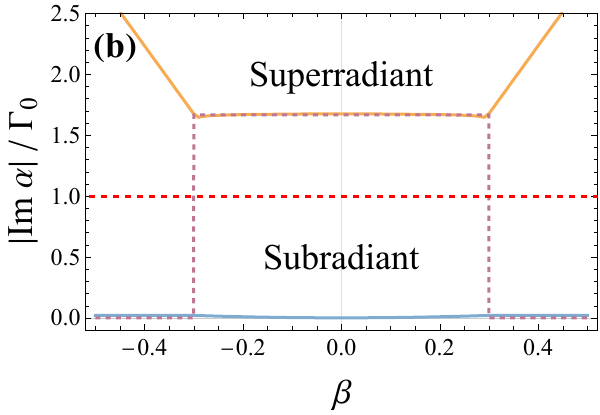}}
   \caption{Band structure of a 1D atomic lattice. Solid lines show the numerical solution for the lower (subradiant) and upper (superradiant)branches of the band structure, while dashed lines show the result from the pole approximation (Eq.~\eqref{eq:pole approx}). (a) Real part of the complex energy $\alpha(\beta)$, corresponding to the collective frequency shift. (b) Imaginary part of the energy, representing the collective decay rate $\Gamma(\beta)$. The horizontal red dashed line indicates the single-atom spontaneous emission rate $\Gamma_0$ for comparison. We set $\alpha_{0} = 0.30$ and $\kappa = 5 \times 10^{-3}$ here and throughout.}
   \label{fig:1d band structure}
\end{figure}

We first analyze the one-dimensional atomic lattice. In this case, the lattice sum can be expressed in closed form (see Appendix~\ref{appd:1d lattice sum}):
\begin{equation}
    \begin{split}
       \mathrm{S}^{(1)}(\alpha,\beta) = & \displaystyle{B}_{2}(\left\vert \beta \right\vert)  -2\alpha^{2}+2\alpha^{2} \ln \alpha + \alpha \ln (2 \pi (\alpha+\left\vert \beta \right\vert ))-\alpha \ln \Gamma(1+\alpha+   \beta  )\\
       &-\alpha \ln \Gamma(1+\alpha-  \beta )-\alpha \ln (1-{e}^{2 \pi \mathrm{i} (\alpha +  \beta  )})-\alpha \ln (1-{e}^{2 \pi \mathrm{i} (\alpha - \beta )}) ,
    \end{split}
    \label{eq:result of 1d lattice sum}    
\end{equation}
where $B_2$ is the second Bernoulli polynomial.
Substituting this expression into Eq.~\eqref{eq: pole equation}, we then solve for the roots $\alpha(\beta)$ in the FBZ. 
%
% HOW ?
%
The resulting complex band structure is displayed in Fig.~\ref{fig:1d band structure}. Throughout this paper, we set $\alpha_{0}=0.30$ and $\kappa = 5 \times 10^{-3}$.
%
% WHY?
%

Figures~\ref{fig:1d band structure}(a) and (b) depict the real and imaginary components of the band energy, revealing a polaritonic structure with distinct upper and lower branches separated by a gap \cite{deutsch1995photonic,sheremet2023waveguide}. We compare the decay rate $\mathrm{Im}(\alpha(\beta))$ to the single-atom rate of spontaneous emission $\Gamma_{0}=2\pi \kappa \alpha_{0}^{2}$ (red dashed line). 
%
% DEFINE \Gamma_0
%
The upper branch is superradiant and exhibits fast decay, while the lower branch is subradiant with slow decay. Notably, a significant portion of the lower band possesses a vanishing imaginary part, corresponding to non-radiative dark states.

To elucidate the physics near resonance ($\alpha \approx \alpha_{0}$), we employ the pole approximation. This corresponds to a perturbative solution of Eq.~\eqref{eq: pole equation}, which holds in the weak coupling regime ($\kappa \ll 1$). 
%
% MORE HERE
%
To first order in $\kappa$ we find that
\begin{equation}
    \alpha(\beta) = \alpha_{0} - 2 \pi \mathrm{i} \kappa \alpha_{0}^{2} - \kappa \mathrm{S}^{(1)}(\alpha_{0},\beta)+ O(\kappa^{2}).
    \label{eq:pole approx}
\end{equation}
As demonstrated by the dashed lines in Fig.~\ref{fig:1d band structure}, this approximation successfully captures the essential features of the band structure near  resonance---both the real and imaginary parts of the band structure in the pole approximation are close to the numerically obtained solutions. We note that at resonance  ($\left\vert \beta \right\vert = \alpha_{0}$), the real part of the energy exhibits a sharp dip, while the imaginary part is significantly suppressed. This suppression arises because modes with $\left\vert \beta \right\vert > \alpha_{0}$ lie outside of the light cone ($|q| > \omega/c$) and are nonradiative.

%===================================================%%

\subsection{\label{subsec:decay rate}Decay Rate}
We now examine the dependence of the collective decay rate $\Gamma(\beta) = |\mathrm{Im}(\alpha(\beta))|$ on the lattice spacing. We find that within the accuracy of the pole approximation,
\begin{equation}
    \Gamma(\beta) = 2\pi \kappa \alpha_{0}^{2}+\pi \kappa \alpha_{0}  \mathrm{sgn} (\sin( 2 \pi (\alpha_{0}+\beta))) \left\vert (\alpha_{0}+\beta)\mathrm{mod} 1-\frac{1}{2} \right\vert + \left( \beta \to  -\beta \right),
    \label{eq:1d decay rate formula}
\end{equation}
where the first term represents the contribution from the spontaneous decay of a single atom, which scales quadratically with $\alpha_{0} = \Omega a / 2 \pi c$. The second term, which scales linearly with $\alpha_{0}$, characterizes the decay rate of the  lattice modes. Eq.~\eqref{eq:1d decay rate formula} can be validated in two asymptotic limits. As $a \to \infty$ ($\alpha_{0} \to \infty$), the quadratic term is dominant, and $\Gamma$ converges to the single-atom rate $\Gamma_{0}$. In addition, as $a \to 0$ ($\alpha_{0} \to 0$), the zero-momentum mode ($\beta=0$) becomes  superradiant ($\Gamma / \Gamma_{0} \to \infty$), while all other modes become subradiant ($\Gamma / \Gamma_{0}  \to 0$).

\begin{figure}
    \subfigure{\includegraphics[width=0.45\textwidth]{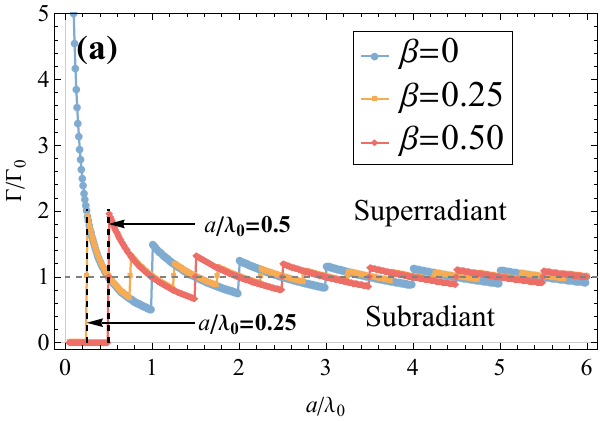}}
    \subfigure{\includegraphics[width=0.45\textwidth]{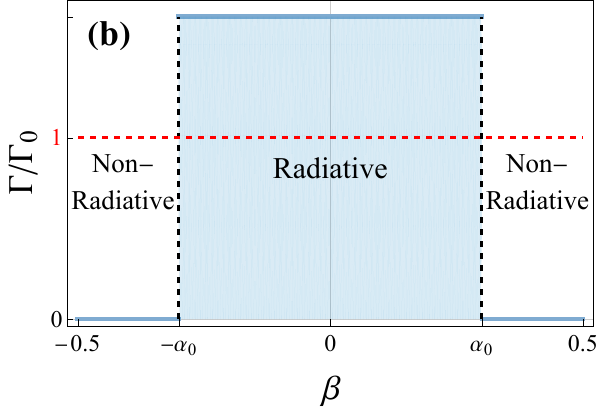}}
    \caption{Decay rate versus lattice spacing in a 1D lattice. (a) The collective decay rate $\Gamma(\beta)$ (normalized by the single-atom rate $\Gamma_0$) calculated via the pole approximation as a function of the dimensionless lattice spacing $\alpha_0$ for various dimensionless momenta $\beta$. The curves display strong oscillations alternating between subradiant ($\Gamma/\Gamma_0<1$) and superradiant ($\Gamma/\Gamma_0>1$) behavior. (b) Schematic of the first Brillouin zone illustrating the light-cone boundary $|\beta|=\alpha_0$: modes are non-radiative for $|\beta|>\alpha_0$ and become radiative when $\alpha_0>| \beta|$. The vertical dashed lines in (a) indicate the critical spacings $\alpha_0=|\beta|$ where each mode crosses the light-cone and radiative decay is activated.}
    \label{fig:1d decay rate}
\end{figure}

Figure~\ref{fig:1d decay rate} illustrates the decay rate as a function of the dimensionless spacing $\alpha_0$. The rate exhibits  non-monotonic behavior, characterized by oscillations that alternate between superradiant ($\Gamma / \Gamma_{0} > 1$) and subradiant ($\Gamma / \Gamma_{0} < 1$) regimes. This result is in quantitative agreement with the numerical observations reported in Ref.~\cite{sierra2022dicke} (see Fig.~1(b) therein), confirming that Eq.~\eqref{eq:1d decay rate formula} captures the underlying physics of the problem.
%
% DO WE USE THE SAME PARAMETERS?
%

The oscillations are explained by two mechanisms. First, the phase factor in the inter-atomic interactions induces interferences that range from constructive to destructive as the distance varies. Second, the radiative coupling is modulated by the light-cone boundary, as depicted in Fig.~\ref{fig:1d decay rate}(b). 
%
% I DON'T UNDERSTAND THIS
%
The mode labeled by $\beta$ remains non-radiative until the spacing increases sufficiently so that $\alpha_0 > |\beta|$, triggering the onset of radiative decay. 
%The interplay between the continuous evolution of the phase and the discrete entry of modes into the radiative region generates the complex oscillatory pattern.

%===================================================%

\subsection{\label{subsec:1d dynamics}Dynamics}
\begin{figure}
    \centering
    \subfigure{\includegraphics[width = 0.435\linewidth]{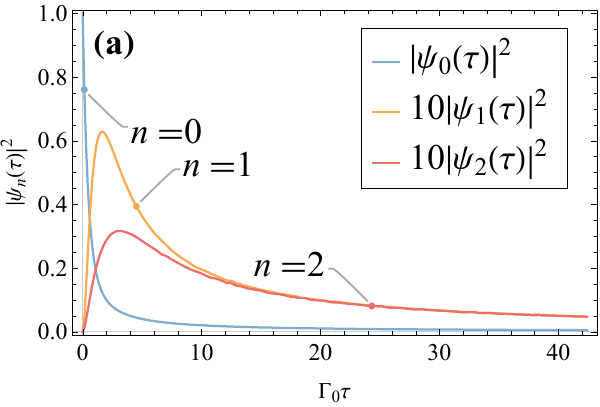}}
    \subfigure{\includegraphics[width = 0.45\linewidth]{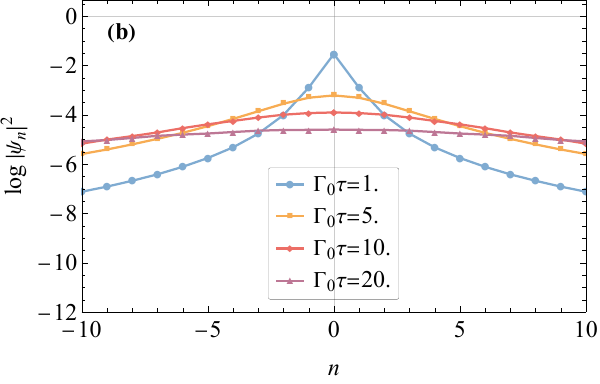}}
    \caption{Illustrating the dynamics in a 1D lattice. (a) Temporal evolution of the atomic population $|\psi_n(t)|^2$ for the initially excited atom ($n=0$) and its nearest neighbors ($n=1, 2$), showing excitation transfer and decay with dimensionless time $\tau = c t / a $. (b) The spatial profile of the excitation probability $|\psi_n(t)|^{2}$ at different time snapshots, illustrating the formation of a localized, non-propagating wavepacket.}
    \label{fig:1D atomic dynamics}
\end{figure}

Finally, we analyze the dynamics of the atomic system. We suppose that only the atom at the origin is excited at $t=0$ so that $\psi(\bm{\beta},0) = 1$. Making use of the pole approximation and computing the inverse Fourier transform of $\psi(\mathbf{q},t)$ defined by Eq.~\eqref{eq:inv_laplace} yields the spatial dependence of the dynamics.
%
% MORE HERE
%
%We initialize the system with the atom at the origin excited at $t=0$ ($\psi_{n}(0) = \delta_{\mathbf{r}_{n}, \mathbf{0}}$), implying uniform population in momentum space ($\psi(\bm{\beta},0) = 1$). We obtain the dynamics via inverse transforms using the analytical pole approximation.
Fig.~\ref{fig:1D atomic dynamics}(a) shows the evolution of $|\psi_n(t)|^{2}$ for the excited atom ($n=0$) and its nearest neighbors ($n=1, 2$). The initial state decays nearly exponentially, transferring its excitation to adjacent sites, which rise transiently before radiating energy to the far field. The spatial dependence of $\psi$, as illustrated in
Fig.~\ref{fig:1D atomic dynamics}(b), reveals that the excitation remains strongly localized. The amplitude peaks at the center and decays rapidly with distance. Rather, it forms a localized wavepacket that dissipates energy into the far-field faster than interatomic transport can occur, thereby inhibiting long-range propagation. 
% I DON'T UNDERSTAND THIS

%%===================================================%%
%%           Two Dimension                           %%
%%===================================================%%
\section{\label{sec:2d}Two Dimensional Lattice}

%%===================================================%%

\subsection{\label{subsec:2d band structure}Band Structure}
\begin{figure}
   \subfigure{\includegraphics[width = 0.45\textwidth]{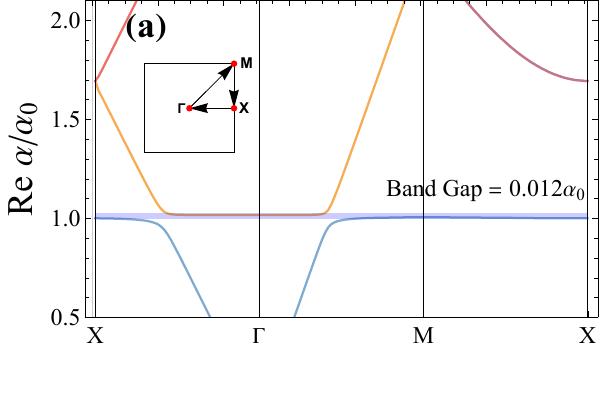}} 
   \subfigure{\includegraphics[width = 0.445\textwidth]{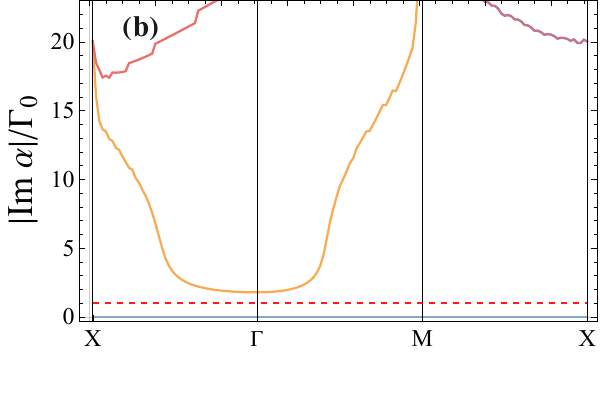}} 
   \caption{Band structure of a 2D square atomic lattice. (a) Real part of the complex energy, showing the polaritonic dispersion. The shaded region highlights the band gap. (b) Imaginary part of the energy, representing the collective decay rate. Note the significant superradiant enhancement at the high-symmetry points.}
   \label{fig:2D band structure}
\end{figure}

We now consider the case of a 2D square lattice of atoms. Unlike the 1D case, the lattice sum does not possess a simple closed form. However, we can derive a rapidly convergent analytical expression for the sum using the theta function transform and Ewald summation (see Appendix~\ref{appd:d-dimensional lattice sum} for details), which is given by:
\begin{equation}
    \begin{aligned}
    \mathrm{S}^{(2)}(\alpha,\bm{\beta}) &= \int_{0}^{1}dt \left[ \frac{1}{2\pi^{2} t }- \frac{\alpha}{2\sqrt{\pi}}\frac{ {e}^{- \frac{\pi^{2} \alpha^{2}}{\ln t }}}{t \sqrt{- \ln t }}  \mathrm{erfc}\left( \frac{\pi \alpha}{\sqrt{- \ln t }} \right) \right] \left[ \prod_{i=1}^{2}\vartheta_{3}(\pi \beta_{i},t) - 1 \right]\\ 
    &+\frac{\alpha}{2} \sum\limits_{\mathbf{n} \neq 0} \left[ {e}^{2 \pi \mathrm{i} \alpha \left\vert \mathbf{n} \right\vert } \mathrm{erfc}(\frac{\left\vert \mathbf{n} \right\vert }{ \eta} + \pi \mathrm{i} \alpha \eta) + {e}^{ - 2 \pi \mathrm{i} \alpha \left\vert \mathbf{n} \right\vert } \mathrm{erfc}(\frac{\left\vert \mathbf{n} \right\vert }{ \eta} - \pi \mathrm{i} \alpha \eta) \right] {e}^{- 2 \pi \mathrm{i} \bm{\beta} \cdot \mathbf{n}} \\
    &+ \alpha \sum\limits_{\mathbf{h}} \frac{\mathrm{erfc}(\pi \eta \sqrt{ \left\vert \bm{\beta} +\mathbf{h} \right\vert^{2} - \alpha^{2} })}{\sqrt{ \left\vert \bm{\beta} +\mathbf{h} \right\vert^{2} - \alpha^{2} } } - \frac{ 2 \alpha }{ \sqrt{\pi} \eta} {e}^{\pi^{2} \alpha^{2} \eta^{2}} - 2 \pi \mathrm{i} \alpha^{2} \mathrm{erfc}(- \pi \mathrm{i} \alpha \eta),
    \end{aligned}
    \label{eq: result of 2d lattice sum}
\end{equation}
where $\eta$ is a free parameter that arises in Ewald summation. Numerical solutions of the Eq.~\eqref{eq: pole equation} yield the band structure presented in Fig.~\ref{fig:2D band structure}, which is displayed at points of high symmetry in the FBZ.
%
% HOW IS THE EQUATION SOLVED?
%

The 2D band structure shares certain features with the 1D case, most notably the existence of a gap between the first and second bands. However, a distinct difference emerges at the high-symmetry points $\Gamma$, X, and M. As illustrated in Fig.~\ref{fig:2D band structure}(b), the imaginary part of the energy is dramatically enhanced at these points, indicating the presence of highly superradiant states. This enhancement is a consequence of Bragg resonance, 
%
% where are the Bragg resonances?
%
where atomic emissions interfere constructively. Notably, the superradiant decay rates in the 2D lattice significantly exceed those found in the 1D case. This amplification arises from the higher dimensionality: an atom in a 2D array possesses a greater number of neighbors at any given distance compared to a 1D chain, thereby facilitating stronger cooperativity and photon emission. Conversely, the lowest energy band exhibits a continuum of long-lived subradiant states, mirroring the behavior observed in 1D.

%===================================================%

\subsection{\label{subsec:2d decay rate}Decay Rate}
\begin{figure}
    \subfigure{\includegraphics[width = 0.55\textwidth]{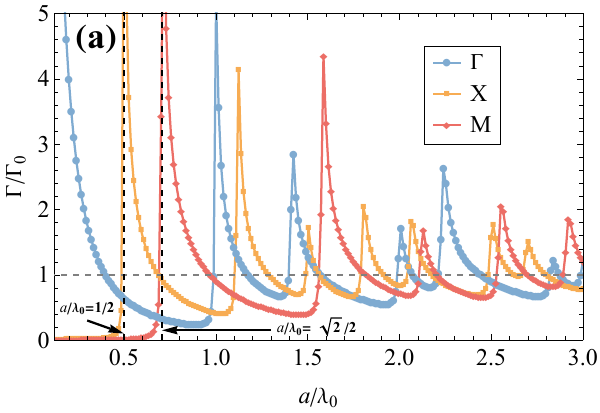}}
    \subfigure{\includegraphics[width = 0.4\textwidth]{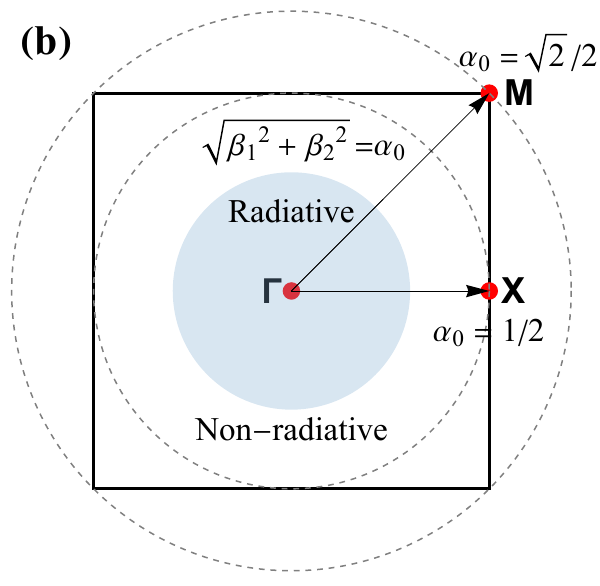}} 
    \caption{Decay rate versus lattice spacing in a 2D square lattice. (a) Collective decay rate $\Gamma(\bm{\beta})$ as a function of the dimensionless lattice spacing $\alpha_0=a/\lambda_0$ for the high-symmetry points $\Gamma$, X, and M, where $\lambda_0=$. The vertical dashed lines indicate the critical thresholds $\alpha_0 = 1/2$ and $\alpha_0 = \sqrt{2}/2$, below which the X and M points, respectively, become non-radiative dark states. (b) Illustration of the light cone in the first Brillouin zone. Modes located inside the circular region, defined by the boundary $\alpha(\bm{\beta}) = \sqrt{\beta_{1}^{2}+\beta_{2}^{2}}$, are radiative, while those outside are non-radiative.}
    \label{fig:2d decay rate}
\end{figure}

We now investigate the collective decay rate of the system, employing the pole approximation used in 1D. Fig.~\ref{fig:2d decay rate}(a) displays the decay rate as a function of lattice spacing for the $\Gamma$, X, and M points. Similar to the 1D results, the rate oscillates with lattice spacing, converging to the single-atom rate $\Gamma_{0}$ as $a \to \infty$ and diverging for small spacings at the $\Gamma$ point due to the break down of the light cone.

The radiative coupling is governed by the light-cone condition $|\bm{\beta}| < \alpha_0$, as depicted in Fig.~\ref{fig:2d decay rate}(b). As the lattice spacing $a$ decreases, the radius of the light cone $\alpha_0 = a/\lambda_0$ decreases, where $\lambda_0=2\pi c/\Omega$. Nevertheless, the $\Gamma$ point ($\bm{\beta}=0$) remains radiative for all spacings. However, the X and M points only become radiative when they fall within the light cone, which occurs at the thresholds $\alpha_0=1/2$ and $\alpha_0=\sqrt{2}/2$, respectively. Below these critical values (marked by vertical dashed lines in Fig.~\ref{fig:2d decay rate}(a)), the corresponding modes can be regarded as non-radiative dark states.

The oscillatory behavior at the $\Gamma$ point (dotted blue line in Fig.~\ref{fig:2d decay rate}(a)) shows good quantitative agreement with numerical simulations reported in Ref.~\cite{rui2020subradiant} (see Fig.~2d, yellow solid line therein). In particular, our calculations accurately reproduce the reported peak positions. The first peak at $a / \lambda_{0} = 1$ arises because the atoms are in phase with their nearest neighbors, leading to maximum constructive interference. Similarly, the second peak at $a / \lambda_{0} = \sqrt{2}$ corresponds to atoms being in phase with their next-nearest neighbors, where the phase difference is $\sqrt{2} a k_{0}  = 2 \lambda_{0}  k_{0} = 4 \pi$. Furthermore, both our results and the numerical simulations show transitions between superradiant and subradiant regimes at approximately the same lattice spacings ($a / \lambda_{0} \approx 0.4, 1.0, 1.2, 1.4, 1.6$). 
%Thus, Eq.~\eqref{eq: result of 2d lattice sum} serves as a reasonable analytical framework for explaining the oscillatory decay rates in 2D atomic lattices.

%===================================================%

\subsection{\label{subsec:2d dynamics}Dynamics}
\begin{figure}
    \centering
    \subfigure{\includegraphics[width = 0.45\textwidth]{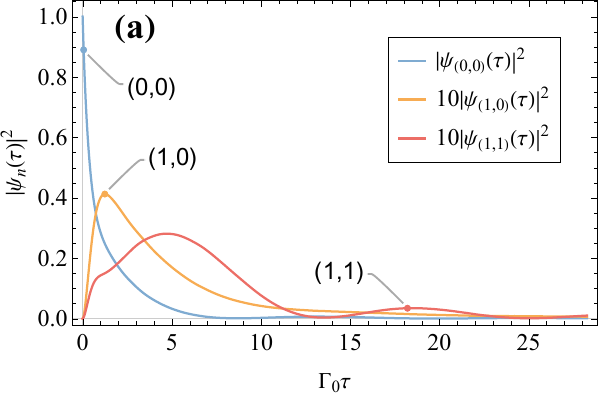}}
    \subfigure{\includegraphics[width = 0.4\textwidth]{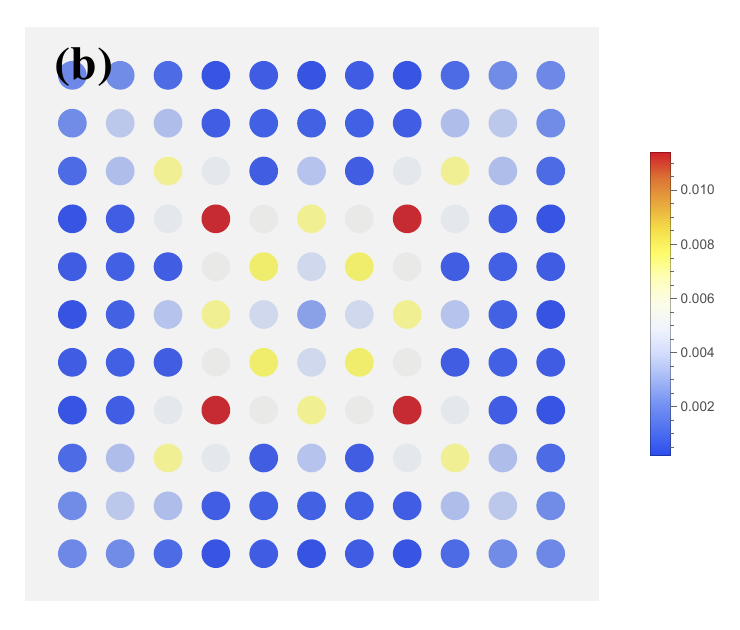}}
    \caption{Spatiotemporal dynamics in a 2D lattice. (a) Temporal evolution of the atomic population $|\psi_n(t)|^2$ for the central atom (0,0) and its neighbors (1,0) and (1,1). (b) Spatial snapshot of the excitation probability $|\psi_n(t)|^{2}$ at time $\Gamma_{0} \tau = 10.0$, showing the outward propagation.}
    \label{fig:2D atomic dynamics}
\end{figure}

We now analyze the dynamics of a single excitation that is initially located at the origin.
The temporal evolution of the populations for the atom and its neighbors, shown in Fig.~\ref{fig:2D atomic dynamics}(a), resembles the 1D case. The excitation transfers to adjacent sites before eventually decaying. However, the spatial profile in Fig.~\ref{fig:2D atomic dynamics}(b) reveals a distinctly different behavior. Unlike the localized wavepacket observed in 1D, the 2D excitation propagates significantly further from the origin. This difference arises from the system geometry: photons can only escape the lattice into free space (the third dimension), whereas in the two planar dimensions, emitted photons are likely to be recaptured by other atoms. Consequently, the rate of in-plane coherent transport competes effectively with out-of-plane radiative decay, allowing the wavepacket to travel a discernible distance before it is dissipated.

%%===================================================%%
%%           Three Dimensions                        %%
%%===================================================%%

\section{\label{sec:3d}Three Dimensional Lattice}
\subsection{\label{subsec:3d band structure}Band Structure}
\begin{figure}
   \includegraphics[width = 0.91\textwidth]{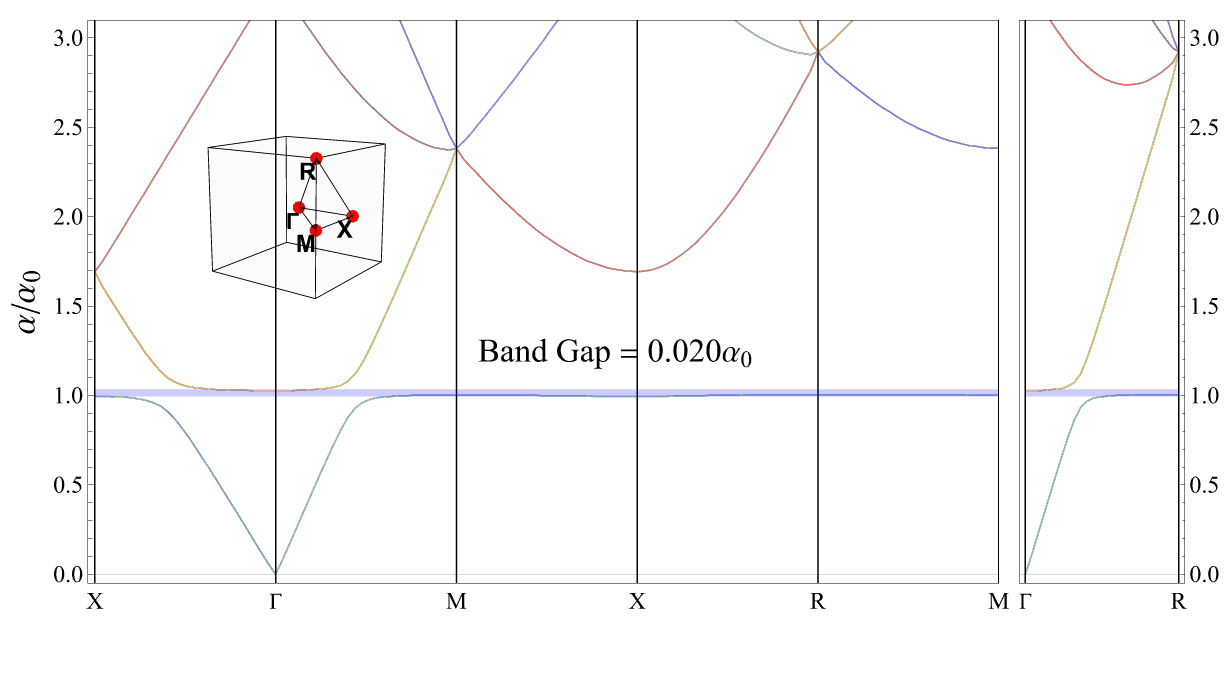}
   \caption{Band structure of a 3D simple-cubic atomic lattice. The bands shown are purely real, indicating that the collective excitations are non-radiative. A band gap opens between the first and second bands as indicated by the shaded region.}
   \label{fig:3D band structure}
\end{figure}

We consider the case of a three-dimensional simple-cubic lattice of atoms. Applying the theta-function transform yields the following expression for the lattice sum:
\begin{equation}
    \begin{aligned}
    \mathrm{S}^{(3)}(\alpha,\bm{\beta}) &= \int_{0}^{1}dt \left[ \frac{1}{2\pi^{2} t }- \frac{\alpha}{2\sqrt{\pi}}\frac{ {e}^{- \frac{\pi^{2} \alpha^{2}}{\ln t }}}{t \sqrt{- \ln t }}  \mathrm{erfc}\left( \frac{\pi \alpha}{\sqrt{- \ln t }} \right) \right] \left[ \prod_{i=1}^{3}\vartheta_{3}(\pi \beta_{i},t) - 1 \right]\\
    &+\frac{\alpha}{2} \sum\limits_{\mathbf{n} \neq 0} \left[ {e}^{2 \pi \mathrm{i} \alpha \left\vert \mathbf{n} \right\vert } \mathrm{erfc}(\frac{\left\vert \mathbf{n} \right\vert }{ \eta} + \pi \mathrm{i} \alpha \eta) + {e}^{ - 2 \pi \mathrm{i} \alpha \left\vert \mathbf{n} \right\vert } \mathrm{erfc}(\frac{\left\vert \mathbf{n} \right\vert }{ \eta} - \pi \mathrm{i} \alpha \eta) \right] {e}^{- 2 \pi \mathrm{i} \bm{\beta} \cdot \mathbf{n}} \\
    &+ \frac{\alpha}{\pi} \sum\limits_{\mathbf{h}}  \frac{ {e}^{- \pi^{2} \eta^{2} \left( \left\vert \bm{\beta} +\mathbf{h} \right\vert^{2} -\alpha^{2}  \right)}}{  \left\vert \bm{\beta} +\mathbf{h} \right\vert^{2} -\alpha^{2}   } -  \frac{2 \alpha}{\sqrt{\pi} \eta} {e}^{\pi^{2} \alpha^{2} \eta^{2}} - 2 \pi \mathrm{i} \alpha^{2} \mathrm{erfc}(- \pi \mathrm{i} \alpha \eta).
    \end{aligned}
    \label{eq: 3d lattice sum}
\end{equation}
The resulting band structure, shown in Fig.~\ref{fig:3D band structure}, reveals a fundamental distinction compared to the lower dimensional cases: the energy bands are entirely real. This indicates that collective excitations in 3D are inherently non-radiative (stable). Unlike the case of 1D and 2D lattices, where photons can escape into the transverse non-confined dimensions, this cannot happen in 3D, 
leading to the inhibition of spontaneous emission. This effect, with classical light, is also found in photonic crystals~\cite{yablonovitch1987inhibited}. Despite the lack of decay, a significant gap persists between the first and second bands. We note that this gap is larger than in 1D and 2D systems. Consequently, the collective Lamb shift—the detuning of the real part of the band energy due to interatomic interactions—is significantly stronger.

The non-radiative nature of the bands is further elucidated by analyzing Eq.~\eqref{eq: pole equation} by making use of the Poisson summation formula:
\begin{equation}
    \alpha - \alpha_{0} + \frac{\kappa}{2\pi} \lim_{\epsilon \to 0^{+}} \sum\limits_{\mathbf{h} \in \mathbb{Z}^{3}} \frac{1}{\left\vert \bm{\beta} +\mathbf{h} \right\vert - \alpha - \mathrm{i} \epsilon } = 0.
    \label{eq: pole equation 3D}
\end{equation}
See Appendix \ref{appd:d-dimensional lattice sum}.
The summation over reciprocal lattice vectors $\mathbf{h}$ remains purely real unless the resonance condition $\alpha = \left\vert \bm{\beta} + \mathbf{h} \right\vert$ is satisfied. This condition corresponds to Bragg scattering, where a collective mode matches a propagating free-space photon. Away from these discrete resonances, the solutions are real, corresponding to a non-radiative dark state.

%===================================================%%

\subsection{\label{subsec:3d decay rate}Decay Rate}
\begin{figure}[htbp]
    \centering
    \subfigure{\includegraphics[width = 0.55\textwidth]{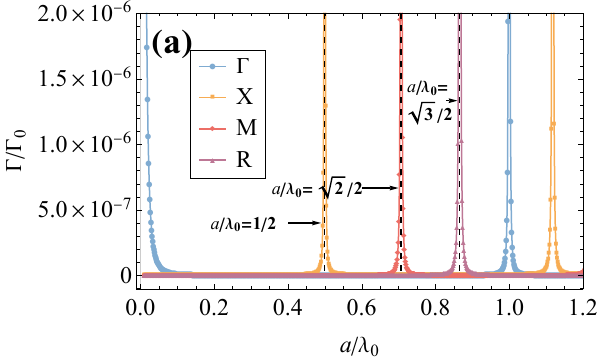}}
    \subfigure{\includegraphics[width = 0.38\textwidth]{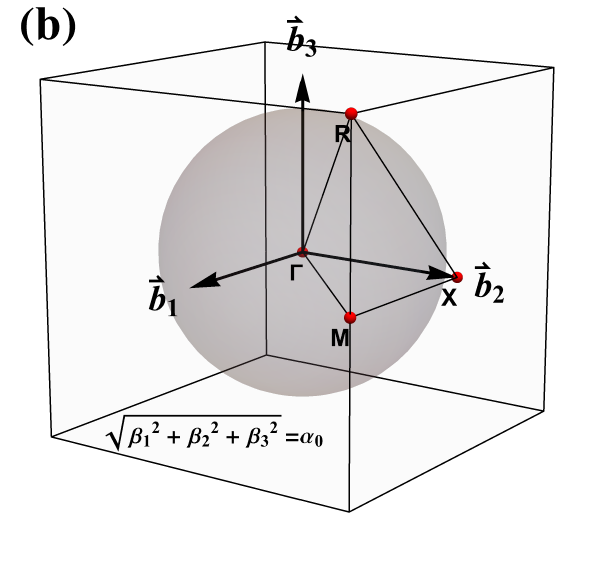}}
    \caption{Resonant decay in a 3D simple-cubic lattice. (a) Collective decay rate $\Gamma(\bm{\beta})$ as a function of the dimensionless lattice spacing $\alpha_0 = a/\lambda_0$ for the high-symmetry points $\Gamma$, X, M, and R. The decay rate is zero except when Bragg resonance occurs, manifested as sharp superradiant peaks. (b) The first Brillouin zone, showing the locations of high-symmetry points. The inscribed sphere indicates the range of momenta accessible to emitted photons.}
    \label{fig:3d decay rate}
\end{figure}

Fig.~\ref{fig:3d decay rate} confirms that the collective decay rate vanishes for the majority of lattice spacings. That is, radiative decay is suppressed except when specific Bragg conditions are met:
\begin{equation}
    a / \lambda_{0} = \frac{m}{2} {C},~m=1,2,3,\ldots,
\end{equation}
where the geometric factor ${C}$ depends on the high-symmetry point (X: $C = 1$, M: $C = \sqrt{2}$, R: $C = \sqrt{3}$). At these frequencies, the lattice functions as a perfectly reflecting mirror. Such resonances are of interest in photonics, especially in high-quality reflectors \cite{deutsch1995photonic, ivchenko1994resonant, corzo2016large, sorensen2016coherent, yanik2004stopping}.

%===================================================%

\subsection{\label{subsec:3d dynamics}Dynamics}
\begin{figure}[htbp]
    \centering
    \subfigure{\includegraphics[width = 0.5\textwidth]{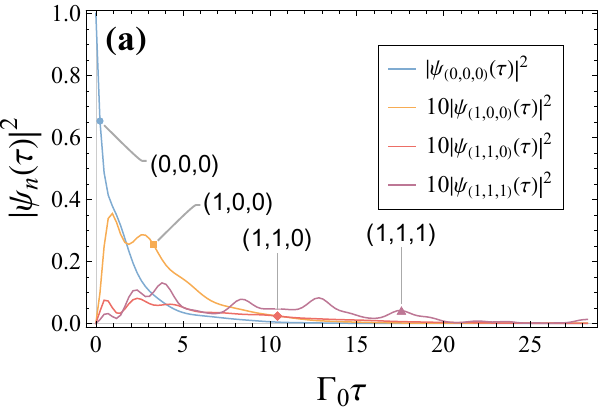}}
    \subfigure{\includegraphics[width = 0.45\textwidth]{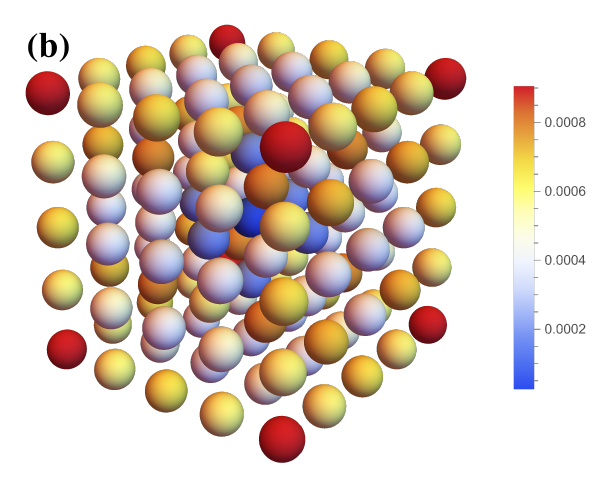}}
    \caption{Dynamics in a 3D lattice. (a) Temporal evolution of the atomic population $|\psi_n(t)|^2$ at selected lattice sites. The observed decay is driven primarily by the coherent spreading of the wavepacket. (b) Spatial profile of the excitation probability $|\psi_n(t)|^{2}$ at time $\Gamma_{0}\tau=10.0$ for a system initialized with a single excitation at the origin. The wavepacket exhibits long-range propagation rather than localized radiative decay.}
    \label{fig:3D atomic dynamics} 
\end{figure}

The dynamics of excitations in the 3D lattice [Fig.~\ref{fig:3D atomic dynamics}] contrasts sharply with its behavior in lower dimensions. The spatial profile shown in Fig.~\ref{fig:3D atomic dynamics}(b) demonstrates that atomic excitations are transported across the entire lattice, which is a direct consequence of the non-radiative band structure, preventing energy loss to the environment. We note that the population $|\psi_n|^2$ is relatively large at substantial distances from the origin.

The temporal evolution of the population at the origin [Fig.~\ref{fig:3D atomic dynamics}(a)] exhibits oscillatory behavior driven by the propagation of excitations rather than radiative loss. The observed reduction in local population is consistent with unitarity: as the wavepacket spreads throughout the lattice, the local amplitude at the origin decreases while the total probability is conserved. This mechanism is distinct from the dissipative exponential decay observed in 1D and 2D. While a minor secondary contribution to decay arises from the small fraction of  states near the Bragg resonances, the dynamics are dominated by non-dissipative coherent transport.

%%===================================================%%
%%           Discussion                              %%
%%===================================================%%

\section{\label{sec:discussion}Discussion}
In summary, we have developed a framework to investigate the collective properties of atomic lattices. Our results underscore the critical role of lattice dimensionality. We find that 1D and 2D systems are inherently radiative, characterized by complex band structures and the coexistence of superradiant and subradiant states. The associated dynamics is dissipative, where excitations either remain localized or decay over short distances before. In contrast, 3D lattices are fundamentally non-radiative, facilitating the formation of photonic band gaps. In this regime, photon emission is strongly inhibited, and radiative decay is restricted to discrete Bragg modes, resulting in dynamics dominated by coherent transport.

The present study employed a scalar model of the electromagnetic field and simple lattice geometries, our framework is readily adaptable to more complex systems. Extension to the full vector electromagnetic field requires replacing the scalar Green's function with its tensor counterpart \cite{PhysRevA.96.063801,chew1999waves,PhysRevA.57.3931}. We emphasize that the theta function transformation and Ewald summation techniques can be adapted to handle the near-field contribution to the tensor Green's function.
Similarly, our results  can be generalized to other lattices, including hexagonal or face-centered cubic structures.

Future research directions include calculating the optical response of atomic lattices to reveal key single-photon properties, such as the transmittance and reflectance of the system \cite{rui2020subradiant}. It would also be valuable to investigate the impact of disorder on the robustness of photonic band gaps \cite{antezza2009spectrum,PhysRevA.96.053804}. Furthermore, topological aspects of the non-Hermitian nature of the effective Hamiltonian should be further explored. Specific avenues include the study of long-lived edge states and the bulk-boundary correspondence \cite{bettles2017topological,perczel2017topological,perczel2020topological,PhysRevLett.121.086803,PhysRevLett.121.136802,ashida2020non}, which hold promise for applications in quantum information processing \cite{PhysRevA.96.041603} and quantum networking \cite{RevModPhys.83.33}.

\section*{ACKNOWLEDGMENTS}
We thank P. Wang, P. de Maat, and T. Hong for helpful discussions. 

\textit{Data availability}: The data and codes are available from the authors upon reasonable request.

%%===================================================%%
%%           Appendix                                %%
%%===================================================%%
\appendix

%%===================================================%%
%%             Green's Function                      %%
%%===================================================%%

\section{\label{appd:Greens function} Green's Function}
Here we reproduce the derivation of the Green's function for the non-local operator $\sqrt{-\Delta} - k$, originally discussed in Ref.~\cite{hiltunen2024nonlocal}. The Green's function $G(\mathbf{r};k)$ satisfies the equation
\begin{equation}
    \left(\sqrt{-\Delta}-k \right)G(\mathbf{r};k) = \delta(\mathbf{r}).
\end{equation}
an can be expressed as a Fourier integral:
\begin{equation}
    G(\mathbf{r};k) = \frac{1}{(2\pi)^{3}} \int d^{3} {q} \frac{{e}^{\mathrm{i} \mathbf{q} \cdot \mathbf{r}}}{\left\vert  \mathbf{q}\right\vert -k}.
\end{equation}
Assuming $\mathrm{Re} k > 0 $, we utilize the algebraic identity:
\begin{equation}
    \frac{1}{|\mathbf{q}|-k}=\frac{1}{|\mathbf{q}|+k} + \frac{2 k}{|\mathbf{q}|^2-k^2},
\end{equation}
to decompose the Green's function into two distinct components:
\begin{equation}
    G(\mathbf{r};k)=G(\mathbf{r};-k)+2 k G_{\text {helm }}(\mathbf{r};k).
\end{equation}
Here, $G_{\text {helm }}$ represents the standard Green's function for the Helmholtz equation. In three dimensions, subject to the radiation condition, it is given by
\begin{equation}
    G_{\text {helm }}(\mathbf{r};k) = \frac{{e}^{\pm \mathrm{i} k r }}{4\pi r },
\end{equation}
where the sign is chosen to match the sign of $\mathrm{Im}k$ to ensure proper decay at infinity. Consequently, the problem reduces to computing $G(\mathbf{r};-k)$. Using the identity
\begin{equation}
    \frac{1}{|\mathbf{q}|+k}=\frac{-k}{|\mathbf{q}|(|\mathbf{q}|+k)}+\frac{1}{|\mathbf{q}|},
\end{equation}
we write $G(\mathbf{r};-k)$ as:
\begin{equation}
    G(\mathbf{r};-k) = \frac{1}{(2\pi)^{3}} \int d^{3} {q} \frac{{e}^{\mathrm{i} \mathbf{q} \cdot \mathbf{r}}}{\left\vert  \mathbf{q} \right\vert } - \frac{k}{(2\pi)^{3}} \int d^{3} {q} \frac{{e}^{\mathrm{i} \mathbf{q} \cdot \mathbf{r}}}{|\mathbf{q}|(|\mathbf{q}|+k)}.
    \label{eq:Greens function -k}
\end{equation}
The first integral in Eq.~\eqref{eq:Greens function -k} is evaluated as follows:
\begin{equation}
   G_1(\mathbf{r}) = \lim_{\mu \to  0} \frac{1}{(2\pi)^{3}}\int d^{3} {q} \frac{{e}^{- \mu \left\vert \mathbf{q} \right\vert }}{ \left\vert  \mathbf{q} \right\vert }{e}^{\mathrm{i} \mathbf{q} \cdot \mathbf{r}} = \lim_{\mu\to 0} \frac{1}{2 \pi^{2} (\mu^{2} + r^{2})} = \frac{1}{2\pi^{2} r^{2}}.
   \label{eq:Greens function 1}
\end{equation}
The second integral in Eq.~\eqref{eq:Greens function -k} is given by
\begin{equation}
    G_2(\mathbf{r};k) = \frac{-k}{(2\pi)^{3}} \int d^{3} {q} \frac{{e}^{\mathrm{i} \mathbf{q} \cdot \mathbf{r}}}{|\mathbf{q}|(|\mathbf{q}|+k)} = \frac{\mathrm{i} k}{4 \pi^2 r } \int_0^{\infty} d |\mathbf{q}| \frac{{e}^{\mathrm{i} |\mathbf{q}|r}-{e}^{-\mathrm{i} |\mathbf{q}|r}}{|\mathbf{q}|+k} ,
\end{equation}
%Applying Jordan's lemma and the Cauchy integral formula, we deform the integration contour to the imaginary axis:
%\begin{equation}
%     G_2(\mathbf{r};k) = \frac{\mathrm{i} k}{4 \pi^2r} \left( \int_0^{\infty} d|\mathbf{q}| \frac{{e}^{-|\mathbf{q}| r }}{|\mathbf{q}|-\mathrm{i} k}-\int_0^{\infty} d|\mathbf{q}| \frac{{e}^{-|\mathbf{q}| r }}{|\mathbf{q}|+\mathrm{i} k}\right).
%\end{equation}
which can be expressed in terms of the exponential integral function $\displaystyle{E}_1(z)$:
\begin{equation}
    G_2(\mathbf{r};k) = \frac{\mathrm{i} k}{4 \pi^2 r }\left[{e}^{\mathrm{i} kr} \displaystyle{E}_1(\mathrm{i} kr)-{e}^{-\mathrm{i} kr} \displaystyle{E}_1(-\mathrm{i} kr)\right].
   \label{eq:Greens function 2}
\end{equation}
Substituting Eqs.~\eqref{eq:Greens function 1} and~\eqref{eq:Greens function 2} into Eq.~\eqref{eq:Greens function -k} yields
\begin{equation}
    G(\mathbf{r};-k) = \frac{1}{2 \pi^2r^2}-\frac{\mathrm{i} k}{4 \pi^2 r }\left[{e}^{ \mathrm{i} kr} \displaystyle{E}_1( \mathrm{i} kr)-{e}^{- \mathrm{i} kr} \displaystyle{E}_1(- \mathrm{i} kr)\right].
\end{equation}
Combining this result with the expression for $G_{\text {helm }}(\mathbf{r};k)$, we obtain the Green's function:
\begin{equation}
    G(\mathbf{r};k)= \begin{cases} \frac{1}{2 \pi^2r^2}-\frac{\mathrm{i} k}{4 \pi^2 r}\left[{e}^{\mathrm{i} kr} \displaystyle{E}_1(\mathrm{i} kr)\right.\left.-{e}^{-\mathrm{i} k r} \displaystyle{E}_1(-\mathrm{i} kr)\right]+\frac{k {e}^{ \pm \mathrm{i} k r}}{2 \pi r}, & k > 0, \\ \frac{1}{2 \pi^2r^2}-\frac{\mathrm{i} k}{4 \pi^2 r}\left[{e}^{\mathrm{i} kr} \displaystyle{E}_1(\mathrm{i} kr)\right.\left.-{e}^{-\mathrm{i} k r} \displaystyle{E}_1(-\mathrm{i} kr)\right], & k < 0, \\ \frac{1}{2 \pi^2r^2}, & k=0 .\end{cases}
    \label{eq:Greens function}
\end{equation}
For $k>0$, the sign of the Helmholtz term is chosen to ensure appropriate causal behavior.

%%===================================================%%
%%          One dimensional lattice sum              %%
%%===================================================%%

\section{\label{appd:1d lattice sum} One Dimensional Lattice Sum}
In this section, we derive the closed-form expression for the one-dimensional lattice sum. We decompose the total sum into three terms:
\begin{subequations}
   \begin{align}
    \mathrm{S}_{1}^{(1)}(\alpha,\beta) &= \frac{1}{2 \pi^{2}}\sum \limits_{n \neq 0}\frac{{e}^{-2 \pi \mathrm{i} \beta n}}{ |n|^{2}}, \\
    \mathrm{S}_{2}^{(1)}(\alpha,\beta) &= - \frac{\mathrm{i} \alpha}{2\pi} \sum \limits_{n \neq 0} \frac{{e}^{-2 \pi \mathrm{i} \beta n}}{\left\vert n \right\vert } \left[  {e}^{2 \pi \mathrm{i} \alpha \left\vert n \right\vert } \displaystyle{E}_{1}(2 \pi \mathrm{i} \alpha \left\vert n \right\vert )- {e}^{-2 \pi \mathrm{i} \alpha \left\vert n \right\vert } \displaystyle{E}_{1}(-2 \pi \mathrm{i} \alpha \left\vert n \right\vert ) \right],\label{eq:def s2 1D}\\
    \mathrm{S}_{3}^{(1)}(\alpha,\beta) &= \alpha \sum \limits_{n \neq 0} \frac{ {e}^{-2 \pi \mathrm{i} \beta n}}{ \left\vert  n \right\vert } {e}^{2 \pi \mathrm{i} \alpha \left\vert n \right\vert }.
   \end{align}
\end{subequations}
Here, $n \neq 0$ denotes summation over $n \in \mathbb{Z} \backslash \{0\}$.

\subsection{\texorpdfstring{$\mathrm{S}_{1}^{(1)}(\alpha,\beta)$}{S1(1)(alpha, beta)}}
Recalling the definition of the polylogarithm function $\mathrm{Li}_{n}(z) = \sum_{k=1}^{\infty} z^{k} / k^{n}$, 
$\mathrm{S}_{1}^{(1)}(\alpha,\beta)$ can be written as:
\begin{equation}
    \mathrm{S}_{1}^{(1)}(\alpha,\beta) = \frac{1}{2\pi^{2}}\left[ \mathrm{Li}_{2}({e}^{2 \pi \mathrm{i} \beta})+\mathrm{Li}_{2}({e}^{-2 \pi \mathrm{i} \beta})  \right].
\end{equation}
Using the identity relating polylogarithms to Bernoulli polynomials $\displaystyle{B}_{n}(x)$ \cite{NIST:DLMF}:
\begin{equation}
    \displaystyle{Li}_{n}({e}^{2 \pi \mathrm{i} x})+(-1)^{n}\displaystyle{Li}_{n}({e}^{-2 \pi \mathrm{i} x}) = - \frac{(2 \pi \mathrm{i})^{n}}{n !} \displaystyle{B}_{n}(x), ~ (0<x<1),
\end{equation}
we obtain the simple formula
\begin{equation}
    \mathrm{S}_{1}^{(1)}(\alpha,\beta) = \displaystyle{B}_{2}(\left\vert  \beta \right\vert ),
    \label{eq:result of s1}
\end{equation}
which holds for $ \left\vert \beta \right\vert \le 1$.

\subsection{\texorpdfstring{$\mathrm{S}_{2}^{(1)}(\alpha,\beta)$}{S2(1)(alpha, beta)}}
The exponential integral function admits the integral representation
\begin{equation}
    \displaystyle{E}_{1}(z) = {e}^{-z}\int_{0}^{\infty} d  t \frac{{e}^{-t z}}{t+1}, ~ \mathrm{Re} z\ge 0.
    \label{eq:int rep of E1}
\end{equation}
Substituting Eq.~\eqref{eq:int rep of E1} into Eq.~\eqref{eq:def s2 1D} and exchanging the order of summation and integration, we obtain
\begin{equation}
    \mathrm{S}_{2}^{(1)}(\alpha,\beta) = \frac{\mathrm{i}\alpha}{2 \pi } \int_{0}^{\infty} \frac{d t}{t+1} \left\{\ln\left[ 1- {e}^{-2 \pi \mathrm{i} (\alpha t+\beta)} \right] +\ln\left[ 1- {e}^{-2 \pi \mathrm{i} (\alpha t-\beta)} \right] \right\}+\text{c.c}.
    \label{eq:s2 integral form 1}
\end{equation}
We temporarily assume $\alpha$ is real (generalizing to complex $\alpha$ via analytic continuation) and select the branch cut of $\ln z $ from $0$ to $-\infty$. Eq.~\eqref{eq:s2 integral form 1} thus becomes
\begin{equation}
    \mathrm{S}_{2}^{(1)}(\alpha,\beta) = -\alpha \left[ I(\alpha,\beta) + I(\alpha,-\beta) \right],
    \label{eq:s2 integral form 2}
\end{equation}
where the auxiliary integral $I(\alpha,\beta)$ is defined for $\alpha > 0 $ and $ \left\vert \beta \right\vert \le  \frac{1}{2}$ as
\begin{equation}
    I(\alpha,\beta) = \int_{0}^{\infty} \frac{dt}{t+1} \mathrm{sgn}(\sin(2 \pi (\alpha t +\beta))) \left\vert  ( \alpha t +\beta ~ \mathrm{mod} ~ 1) -\frac{1}{2} \right\vert.
\end{equation}
Evaluating this integral accounting for the contributions due to the sgn function leads to
\begin{equation}
    I(\alpha,\beta)=
    \begin{cases} A(2(\alpha-\beta))+(\frac{1}{2}+\alpha-\beta) \ln\left(\frac{\frac{1}{2}+\alpha - \beta}{\alpha}\right)+\beta - \frac{1}{2}, ~  & \beta \ge  0, \\ A(2(\alpha-\beta))+(\frac{1}{2}+\alpha-\beta) \ln\left(  \frac{\frac{1}{2}+\alpha - \beta}{\alpha-\beta}\right)+(-\frac{1}{2}+\alpha-\beta)\ln\left( \frac{\alpha-\beta}{\alpha} \right)+\beta - \frac{1}{2},~ &\beta<0,
    \end{cases}
\end{equation}
where $A(x)$ is the series:
\begin{equation}
    A(x) = \frac{1}{2}\sum \limits_{n=1}^{\infty} \left\{ (2n+1+x) \ln\left( 1+ \frac{1}{2 n + x } \right)+\left( 2n-1+x \right)  \ln \left( 1 + \frac{1}{2 n - 1 + x } \right) -2 \right\}.
    \label{eq:def of A(x)}
\end{equation}
This series is absolutely convergent, enabling us to interchange the differentiation and summation. We observe that
\begin{equation}
     \frac{d^{2}}{dx^{2}}A(x) = - \sum \limits_{n=1}^{\infty} \frac{1}{(2 n-1+x) (2 n+x)^2 (2 n+1+x)} = \frac{1}{4} \left[ - \frac{2}{1+x}+ \psi^{(1)} (1 + \frac{x}{2})\right],
\end{equation}
where $\psi^{(1)}(z)$ is the polygamma function of order 1. Integrating twice with respect to $x$ recovers $A(x)$ up to constants:
\begin{equation}
    A(x) = -\frac{1}{2} (x+1) \ln (x+1)+{\ln \Gamma  }\left(1+\frac{x}{2}\right)+ C_{1} x+C_{0}.
    \label{eq:undetermined form of A(x)}
\end{equation}
We determine the constants $C_{0}$ and $C_{1}$ by evaluating the limits of the series Eq.~\eqref{eq:def of A(x)} at $x=0$ and $x=1$ using Stirling's formula. This yields $C_{0} = \frac{1}{2}(1-\ln \pi)$ and $C_{1} = \frac{1}{2}(1+\ln 2)$. We thus obtain
\begin{equation}
    A(x) =\frac{1}{2} \left[x+(x+1)\ln\left( \frac{2}{x+1} \right)+2 \ln \Gamma(1+\frac{x}{2})+1-\ln(2\pi) \right].
    \label{eq:final form of A(x)}
\end{equation}
Finally, substituting Eq.~\eqref{eq:final form of A(x)} into Eq.~\eqref{eq:s2 integral form 2}, we arrive at:
\begin{equation}
    \mathrm{S}_{2}^{(1)}(\alpha,\beta)=-2\alpha^{2}+2\alpha^{2} \ln \alpha + \alpha \ln(2\pi(\alpha+\left\vert \beta \right\vert ))-\alpha \ln \Gamma(1+\alpha+\beta)-\alpha \ln \Gamma (1+\alpha-\beta).
    \label{eq:result of s2}
\end{equation}
%We have numerically verified that this result holds for complex $\alpha$ via analytical continuation.

\subsection{\texorpdfstring{$\mathrm{S}_{3}^{(1)}(\alpha,\beta)$}{S3(1)(alpha, beta)}}
$\mathrm{S}_{3}^{(1)}(\alpha,\beta)$ can be evaluated directly:
\begin{align}
    \mathrm{S}_{3}^{(1)}(\alpha,\beta) 
    & =  - \alpha \left[ \ln(1-{e}^{2 \pi \mathrm{i} (\alpha+\beta)}) +\ln(1-{e}^{2 \pi \mathrm{i} (\alpha-\beta)}) \right].
    \label{eq:result of s3}
\end{align}
Crucially, $\mathrm{S}_{3}^{(1)}(\alpha,\beta)$ is complex-valued, arising from the imaginary part of the Helmholtz Green's function, and is responsible for the radiative decay rate of the lattice.

\subsection{Final result of one-dimensional lattice sum}
Combining Eqs.~\eqref{eq:result of s1}, \eqref{eq:result of s2}, and \eqref{eq:result of s3}, the total one-dimensional lattice sum is:
\begin{equation}
    \begin{split}
       \mathrm{S}^{(1)}(\alpha,\beta) = & \displaystyle{B}_{2}(\left\vert \beta \right\vert)  -2\alpha^{2}+2\alpha^{2} \ln \alpha + \alpha \ln (2 \pi (\alpha+\left\vert \beta \right\vert ))-\alpha \ln \Gamma(1+\alpha+   \beta  )\\
       &-\alpha \ln \Gamma(1+\alpha-  \beta )-\alpha \ln (1-{e}^{2 \pi \mathrm{i} (\alpha +  \beta  )})-\alpha \ln (1-{e}^{2 \pi \mathrm{i} (\alpha - \beta )}).
    \end{split}  
\end{equation}
This expression is valid for complex $\alpha$ and $\beta$ within the FBZ. For $\beta$ outside the FBZ, the periodicity relation $\mathrm{S}^{(1)}(\alpha,\beta) = \mathrm{S}^{(1)}(\alpha,\beta+m)$ holds.

%%===================================================%%
%%          Higher dimensional lattice sum           %%
%%===================================================%%

\section{\label{appd:d-dimensional lattice sum}Higher Dimensional Lattice Sums}
We now explain the method for computing the $d$-dimensional lattice sum $\mathrm{S}^{(d)}(\alpha,\bm{\beta})$. As in the 1D case, we split the sum into three parts:
\begin{subequations}
    \begin{align}
       &\mathrm{S}_{1}^{(d)}(\bm{\beta})  = \frac{1}{2 \pi^{2} } \sum_{\mathbf{n} \neq 0} \frac{1}{\left\vert \mathbf{n} \right\vert^{2}} {e}^{-2 \pi \mathrm{i} \bm{\beta} \cdot \mathbf{n}}, \\
       &\mathrm{S}_{2}^{(d)}(\alpha,\bm{\beta}) = -\frac{\mathrm{i} \alpha}{2\pi} \sum_{\mathbf{n} \neq 0} \frac{1}{\left\vert \mathbf{n} \right\vert} \left[ {e}^{2 \pi \mathrm{i} \alpha \left\vert \mathbf{n} \right\vert } \displaystyle{E}_{1}(2 \pi \mathrm{i} \alpha \left\vert \mathbf{n} \right\vert )- {e}^{-2 \pi \mathrm{i} \alpha \left\vert \mathbf{n} \right\vert } \displaystyle{E}_{1}(-2 \pi \mathrm{i} \alpha \left\vert \mathbf{n} \right\vert ) \right] {e}^{-2 \pi \mathrm{i} \bm{\beta} \cdot \mathbf{n}},\\ 
        &\mathrm{S}_{3}^{(d)}(\alpha,\bm{\beta}) = \alpha \sum_{\mathbf{n} \neq 0} \frac{1}{\left\vert \mathbf{n} \right\vert} {e}^{2 \pi \mathrm{i} \alpha \left\vert \mathbf{n} \right\vert } {e}^{-2 \pi \mathrm{i} \bm{\beta} \cdot \mathbf{n}}.
    \end{align}
    \label{eq: def of partial latice sum}
\end{subequations}
where $\mathbf{n} \in  \mathbb{Z}^{d}$. Below, we will employ the theta-function transform and Ewald summation to derive rapidly convergent expressions.

\subsection{\texorpdfstring{$\mathrm{S}_{1}^{(d)}(\bm{\beta})$}{S1(d)(beta)}}
Using the identity $x^{-1} = \int_{0}^{1} dt ~t^{x-1}$ (for $x>0$), $\mathrm{S}_{1}^{(d)}(\bm{\beta})$ can be rewritten as:
\begin{equation}
    \mathrm{S}_{1}^{(d)} = \frac{1}{2 \pi^{2}}  \int_{0}^{1} \frac{dt}{t}  \sum_{\mathbf{n} \neq 0} t^{\mathbf{n}^{2}} {e}^{2 \pi \mathrm{i} \bm{\beta} \cdot \mathbf{n}}.
    \label{eq: S1 step 1}
\end{equation}
Identifying the inner sum with the third Jacobi theta function $\vartheta_{3}(z,q) = \sum_{n=-\infty}^{\infty} q^{n^{2}} {e}^{2  \mathrm{i} n z}$, we obtain:
\begin{equation}
    \mathrm{S}_{1}^{(d)} = \frac{1}{2\pi^{2}} \int_{0}^{1} \frac{dt}{t} \left[\prod_{i=1}^{d} \vartheta_{3} (\pi \beta_{i},t) - 1 \right].
    \label{eq:integral rep of S1}
\end{equation}
While this integral cannot be evaluated in closed-form, it is rapidly convergent and well suited for numerical evaluation.

\subsection{\texorpdfstring{$\mathrm{S}_{2}^{(d)}(\alpha,\bm{\beta})$}{S2(d)(alpha,beta)}}
Using the integral representation of $\displaystyle{E}_{1}(z)$ (Eq.~\eqref{eq:int rep of E1}), $\mathrm{S}_{2}^{(d)}(\alpha,\bm{\beta})$ becomes
\begin{equation}
    \mathrm{S}_{2}^{(d)}(\alpha,\bm{\beta}) = \frac{1}{\pi} \int_{0}^{\infty} \frac{du}{u+1} \sum_{\mathbf{n} \neq 0} \frac{\sin(2 \pi \alpha u \left\vert \mathbf{n} \right\vert )}{\left\vert  \mathbf{n}\right\vert } {e}^{2 \pi \mathrm{i} \bm{\beta} \cdot \mathbf{n}}.
    \label{eq: S2 step 1}
\end{equation}  
We employ the Mellin transform, $\mathcal{M}[f(x)](s) = \int_{0}^{\infty} d x ~ f(x) ~ x^{s-1}$, with respect to $\alpha$ to transform the sum into a product of theta functions. We thereby obtain
\begin{equation}
    \mathcal{M}[\mathrm{S}_{2}^{(d)}(\alpha)](s) = \frac{1}{\pi} \int_{0}^{\infty} \frac{du}{u+1} \sum_{\mathbf{n} \neq 0}(2\pi u )^{-s} \left\vert \mathbf{n} \right\vert^{-(s+1)} \Gamma(s) \sin(\frac{\pi}{2}s) {e}^{2 \pi \mathrm{i} \bm{\beta} \cdot \mathbf{n}}. 
    \label{eq:s2 step 2}
\end{equation}
By using the following integral identities:
\begin{subequations}
    \begin{align}
        \frac{1}{\Gamma(s)} & \int_{0}^{1} d t ~ (- \ln t )^{s-1} t^{x-1} = \frac{1}{x^{s}},~ (\mathrm{Re} s >0), \\
        &\int_{0}^{\infty} \frac{du}{u^{s}(u+1)} = \frac{\pi}{\sin(\pi s )},~ (0 < \mathrm{Re} s <1),
    \end{align}
\end{subequations}
we can rewrite Eq.~\eqref{eq:s2 step 2} in terms of $\vartheta_{3}$:
\begin{equation}
    \mathcal{M}[\mathrm{S}_{2}^{(d)}(\alpha)](s) = \frac{1}{4\sqrt{\pi}}\frac{\pi^{-s}\Gamma(\frac{s}{2})}{\cos(\frac{\pi}{2} s )} \int_{0}^{1} d t ~ \frac{(-\ln t )^{\frac{s-1}{2}}}{t} \left[ \prod_{i=1}^{d} \vartheta_{3}(\pi \beta_{i},t) - 1 \right].
\end{equation}
Inverting the Mellin transform leads to the final expression:
\begin{equation}
    \mathrm{S}_{2}^{(d)}= \frac{1}{2 \sqrt{\pi}} \int_{0}^{1}d t ~ \frac{{e}^ {-\frac{\pi^{2}\alpha^{2}}{\ln t }}}{t\sqrt{- \ln t }} \mathrm{erfc}\left(\frac{\pi \alpha}{\sqrt{- \ln t }} \right) \left[ \prod_{i=1}^{d}\vartheta_{3}(\pi \beta_{i},t) - 1 \right].
    \label{eq:integral rep of S2}
\end{equation}
Numerical implementation of this integral requires care since the integrand contains the rapidly growing function ${e}^{x^{2}} \mathrm{erfc}(x)$. This can be handled using scaled complementary error function routines or continued-fraction expansions \cite{NIST:DLMF, cody1969rational}.

\subsection{\texorpdfstring{$\mathrm{S}_{3}^{(d)}(\alpha,\bm{\beta})$}{S3(d)(alpha,beta)}}
The summand in $\mathrm{S}_{3}^{(d)}$ admits the Ewald integral representation \cite{beylkin2008fast}:
\begin{equation}
    \frac{1}{ \left\vert  \mathbf{n} \right\vert } {e}^{ 2 \pi \mathrm{i} \alpha \left\vert \mathbf{n} \right\vert } = \frac{2}{\sqrt{\pi}} \int_{0{e}^{\mathrm{i} \theta}}^{\infty} du ~ \frac{1}{u^{2}} {e}^{\pi^{2}\alpha^{2} u^{2} - \left\vert \mathbf{n} \right\vert^{2}  / u^{2}},
    \label{eq: Ewald rep}
\end{equation}
where the integration contour angle $\theta$ is chosen to ensure convergence \cite{capolino2007efficient}.  This allows us to write:
\begin{equation}
    \mathrm{S}_{3}^{(d)}(\alpha,\bm{\beta}) = \frac{2 \alpha}{\sqrt{\pi}} \int_{0 {e}^{\mathrm{i} \theta}}^{\infty} \frac{du}{u^{2}} {e}^{\pi^{2} \alpha^{2} u^{2}} \left[ \prod_{i=1}^{d} \vartheta_{3}(\pi \beta_{i},{e}^{ - \frac{1}{u^{2}}}) - 1 \right].
    \label{eq:integral rep of S3}
\end{equation}
Eq.~\eqref{eq:integral rep of S3} is not suited for direct numerical evaluation due to its slow convergence and oscillatory nature. We apply the Ewald summation technique, introducing a parameter $\eta$ to split the integral into a short-range part (real space) and a long-range part (in Fourier space):
\begin{subequations}
\begin{align}
    &\mathrm{S}_{3\text{r}}^{(d)}(\alpha,\bm{\beta};\eta) = \frac{2 \alpha}{ \sqrt{\pi} } \sum\limits_{\mathbf{n} \neq 0} \int_{0{e}^{\mathrm{i} \theta}}^{\eta} du ~ \frac{1}{u^{2}} {e}^{\pi^{2}\alpha^{2} u^{2} - \left\vert \mathbf{n} \right\vert^{2}  / u^{2}} {e}^{- 2 \pi \mathrm{i} \bm{\beta} \cdot \mathbf{n}},\label{eq: def of real space sum} \\
    &\mathrm{S}_{3\text{m}}^{(d)}(\alpha,\bm{\beta};\eta) = \frac{2 \alpha}{ \sqrt{\pi} } \sum\limits_{\mathbf{n} \neq 0} \int_{\eta}^{\infty} du ~ \frac{1}{u^{2}} {e}^{\pi^{2}\alpha^{2} u^{2} - \left\vert \mathbf{n} \right\vert^{2}  / u^{2}} {e}^{- 2 \pi \mathrm{i} \bm{\beta} \cdot \mathbf{n}}. \label{eq: def of momentum space sum}
\end{align}
\label{eq: def of real and momentum space sum}
\end{subequations}
The total sum is independent of $\eta$, which is tuned to optimize computational efficiency (we choose $\eta = 1 / \sqrt{\pi}$). The real-space sum $\mathrm{S}_{3\text{r}}^{(d)}$ integrates to:
\begin{equation}
    \mathrm{S}_{3\text{r}}^{(d)}(\alpha,\bm{\beta};\eta) = \frac{\alpha}{2} \sum\limits_{\mathbf{n} \neq 0} \left[ {e}^{2 \pi \mathrm{i} \alpha \left\vert \mathbf{n} \right\vert } \mathrm{erfc}(\frac{\left\vert \mathbf{n} \right\vert }{ \eta} + \pi \mathrm{i} \alpha \eta) + {e}^{ - 2 \pi \mathrm{i} \alpha \left\vert \mathbf{n} \right\vert } \mathrm{erfc}(\frac{\left\vert \mathbf{n} \right\vert }{ \eta} - \pi \mathrm{i} \alpha \eta) \right] {e}^{- 2 \pi \mathrm{i} \bm{\beta} \cdot \mathbf{n}}.
    \label{eq:real space sum of S3}    
\end{equation}
While for $\mathrm{S}_{3\text{m}}^{(d)}$, we use the Poisson summation formula to obtain
\begin{equation}
   \mathrm{S}_{3\text{m}}^{(d)}(\alpha,\bm{\beta};\eta) = 2 \alpha \pi^{\frac{d-1}{2}} \sum\limits_{\mathbf{h}} \int_{\eta}^{\infty} d u ~ u^{d-2} {e}^{- \pi^{2} \left( \left\vert  \bm{\beta} + \mathbf{h} \right\vert^{2} - \alpha^{2}  \right) u^{2}} - \frac{2 \alpha}{ \sqrt{\pi} } \int_{\eta}^{\infty} du ~ \frac{1}{u^{2}} {e}^{\pi^{2}\alpha^{2} u^{2}}.
   \label{eq:momentum space sum of S3}
\end{equation}
Making use of the incomplete Gamma function $\Gamma(s,x)$
\begin{equation}
    \int_{\eta}^{\infty} d u ~ u^{d-2} {e}^{- x^{2} u^{2}} = \frac{x^{1-d}}{2} \Gamma\left( \frac{d-1}{2}, x^{2} \eta^{2} \right),
    \label{eq:integral identity for S3}
\end{equation}
we can derive explicit expressions for $d=1,2,3$:
\begin{subequations}
   \begin{align}
    &\mathrm{S}_{3\text{m}}^{(1)} = \alpha \sum\limits_{\text{h}} \displaystyle{E}_{1} ( \pi^{2} \eta^{2} ( \left\vert  \beta + \text{h} \right\vert^{2} - \alpha^{2} )) - \frac{ 2 \alpha }{ \sqrt{\pi} \eta} {e}^{\pi^{2} \alpha^{2} \eta^{2}} - 2 \pi \mathrm{i} \alpha^{2} \mathrm{erfc}(- \pi \mathrm{i} \alpha \eta),\\
    &\mathrm{S}_{3\text{m}}^{(2)} = \alpha \sum\limits_{\mathbf{h}} \frac{\mathrm{erfc}(\pi \eta \sqrt{ \left\vert \bm{\beta} +\mathbf{h} \right\vert^{2} - \alpha^{2} })}{\sqrt{ \left\vert \bm{\beta} +\mathbf{h} \right\vert^{2} - \alpha^{2} } } - \frac{ 2 \alpha }{ \sqrt{\pi} \eta} {e}^{\pi^{2} \alpha^{2} \eta^{2}} - 2 \pi \mathrm{i} \alpha^{2} \mathrm{erfc}(- \pi \mathrm{i} \alpha \eta),\\
    &\mathrm{S}_{3\text{m}}^{(3)} = \frac{\alpha}{\pi} \sum\limits_{\mathbf{h}}  \frac{ {e}^{- \pi^{2} \eta^{2} \left( \left\vert \bm{\beta} +\mathbf{h} \right\vert^{2} -\alpha^{2}  \right)}}{  \left\vert \bm{\beta} +\mathbf{h} \right\vert^{2} -\alpha^{2}   } -  \frac{2 \alpha}{\sqrt{\pi} \eta} {e}^{\pi^{2} \alpha^{2} \eta^{2}} - 2 \pi \mathrm{i} \alpha^{2} \mathrm{erfc}(- \pi \mathrm{i} \alpha \eta).
   \end{align} 
   \label{eq:explicit form of S3m}
\end{subequations}

\subsection{Final expression for \texorpdfstring{$d$}{d}-dimensional lattice sum}
Combining Eqs.~\eqref{eq:integral rep of S1}, \eqref{eq:integral rep of S2}, \eqref{eq:real space sum of S3}, and \eqref{eq:explicit form of S3m}, the final expression is:
\begin{equation}
    \begin{aligned}
    \mathrm{S}^{(d)}(\alpha,\bm{\beta}) &=  \int_{0}^{1}dt \left[ \frac{1}{2\pi^{2} t }- \frac{\alpha}{2\sqrt{\pi}}\frac{ {e}^{- \frac{\pi^{2} \alpha^{2}}{\ln t }}}{t \sqrt{- \ln t }}  \mathrm{erfc}\left( \frac{\pi \alpha}{\sqrt{- \ln t }} \right) \right] \left[ \prod_{i=1}^{d}\vartheta_{3}(\pi \beta_{i},t) - 1 \right]\\
    &+\frac{\alpha}{2} \sum\limits_{\mathbf{n} \in \mathbb{Z}^{d}\backslash \{0\}} \left[ {e}^{2 \pi \mathrm{i} \alpha \left\vert \mathbf{n} \right\vert } \mathrm{erfc}(\frac{\left\vert \mathbf{n} \right\vert }{ \eta} + \pi \mathrm{i} \alpha \eta) + {e}^{ - 2 \pi \mathrm{i} \alpha \left\vert \mathbf{n} \right\vert } \mathrm{erfc}(\frac{\left\vert \mathbf{n} \right\vert }{ \eta} - \pi \mathrm{i} \alpha \eta) \right] {e}^{- 2 \pi \mathrm{i} \bm{\beta} \cdot \mathbf{n}} \\
    &+ \alpha  \sum\limits_{\mathbf{h}\in \mathbb{Z}^{d}} \frac{\Gamma\left(\frac{d-1}{2},\pi^{2} \eta^{2}  (\left\vert \bm{\beta} + \mathbf{h} \right\vert^{2} - \alpha^{2}) \right)}{ \left[\pi\left( \left\vert \bm{\beta} + \mathbf{h} \right\vert^{2} - \alpha^{2} \right) \right]^{\frac{d-1}{2}} } -  \frac{2 \alpha}{\sqrt{\pi} \eta} {e}^{\pi^{2} \alpha^{2} \eta^{2}} - 2 \pi \mathrm{i} \alpha^{2} \mathrm{erfc}(- \pi \mathrm{i} \alpha \eta).
    \end{aligned}
\end{equation}

\bibliographystyle{apsrev4-2}
\bibliography{paper}
\end{document}